\documentclass[twocolumn]{aastex7}

\usepackage{enumitem}
\usepackage{amsmath}
\usepackage{latexsym}
\usepackage{graphicx}
\usepackage{epsf}
\usepackage{CJK}
\usepackage{wasysym}
\usepackage{natbib}

\newcommand{\oiii}{[O\,{\sc iii}]}

\newcommand{\nii}{[N\,{\sc ii}]}

\newcommand{\ha}{H$\alpha$}
\newcommand{\hb}{H$\beta$}

\newcommand{\source}{DEEP23-z2LRD1}
\newcommand{\jwst}{\textit{JWST}}

\received{MMDDYY}
\revised{MMDDYY}
\accepted{MMDDYY}

\submitjournal{ApJ}

\shorttitle{Little Red Dots at Cosmic Noon}
\shortauthors{Y. Ma et al}

\graphicspath{{./}{figures/}}

\begin{document}
\begin{CJK*}{UTF8}{gbsn}
\title{Counting Little Red Dots at $z<4$ with Ground-based Surveys and Spectroscopic Follow-up}


\correspondingauthor{Yilun Ma (yilun@princeton.edu)}
\author[0000-0002-0463-9528]{Yilun Ma (马逸伦)}
\affiliation{Department of Astrophysical Sciences, Princeton University, Princeton, NJ 08544, USA}
\email{yilun@princeton.edu}

\author[0000-0002-5612-3427]{Jenny E. Greene}
\affiliation{Department of Astrophysical Sciences, Princeton University, Princeton, NJ 08544, USA}
\email{jgreene@astro.princeton.edu}

\author[0000-0003-4075-7393]{David J. Setton}
\affiliation{Department of Astrophysical Sciences, Princeton University, Princeton, NJ 08544, USA}
\email{davidsetton@princeton.edu}

\author[0000-0003-4700-663X]{Andy D. Goulding}
\affiliation{Department of Astrophysical Sciences, Princeton University, Princeton, NJ 08544, USA}
\email{goulding@astro.princeton.edu}

\author[0000-0002-8053-8040]{Marianna Annunziatella}
\affiliation{Centro de Astrobiolog\'ia (CAB), CSIC-INTA, Ctra. de Ajalvir km 4, Torrej\'on de Ardoz 28850, Madrid, Spain}
\email{mannunziatella@cab.inta-csic.es}

\author[0000-0003-3310-0131]{Xiaohui Fan}
\affiliation{Steward Observatory, University of Arizona, 933 N Cherry Avenue, Tucson, AZ 85721, USA}
\email{xiaohuidominicfan@gmail.com}

\author[0000-0002-5588-9156]{Vasily Kokorev}
\affiliation{Department of Astronomy, The University of Texas at Austin, Austin, TX 78712, USA}
\email{vkokorev@utexas.edu}

\author[0000-0002-2057-5376]{Ivo Labbe}
\affiliation{Centre for Astrophysics and Supercomputing, Swinburne University of Technology, Melbourne, VIC 3122, Australia}
\email{ilabbe@swin.edu.au}

\author[0000-0001-9592-4190]{Jiaxuan Li (李嘉轩)}
\affiliation{Department of Astrophysical Sciences, Princeton University, Princeton, NJ 08544, USA}
\email{jiaxuanl@princeton.edu}

\author[0000-0001-6052-4234]{Xiaojing Lin}
\affiliation{Steward Observatory, University of Arizona, 933 N Cherry Avenue, Tucson, AZ 85721, USA}
\email{xiaojinglin@arizona.edu}

\author[0000-0001-9002-3502]{Danilo Marchesini}
\affiliation{Department of Physics and Astronomy, Tufts University, Medford, MA 02155, USA}
\email{danilo.marchesini@tufts.edu}

\author[0000-0003-2871-127X]{Jorryt Matthee}
\affiliation{Institute of Science and Technology Austria (ISTA), Am Campus 1, 3400 Klosterneuburg, Austria}
\email{jorryt.matthee@ist.ac.at}

\author[0000-0002-6265-2675]{Luke Robbins}
\affiliation{Department of Physics and Astronomy, Tufts University, Medford, MA 02155, USA}
\email{andrew.robbins@tufts.edu}

\author[0000-0002-1917-1200]{Anna Sajina}
\affiliation{Department of Physics and Astronomy, Tufts University, Medford, MA 02155, USA}
\email{annie.sajina@gmail.com}

\author[0000-0002-7712-7857]{Marcin Sawicki}
\affiliation{Institute for Computational Astrophysics and Department of Astronomy and Physics, Saint Mary's University, 923 Robie Street, Halifax, Nova Scotia, B3H 3C3, Canada}
\email{marcin.sawicki@smu.ca}

\author[0000-0003-4122-7749]{O. Grace Telford}
\affiliation{Department of Astrophysical Sciences, Princeton University, Princeton, NJ 08544, USA}
\affiliation{The Observatories of the Carnegie Institution for Science, 813 Santa Barbara Street, Pasadena, CA 91101, USA}
\email{grace.telford@astro.princeton.edu}

\begin{abstract}
Little red dots (LRDs) are a population of red, compact objects discovered by \jwst\ at $z>4$. At $4<z<8$, they are roughly 100 times more abundant than UV-selected quasars. However, their number density is uncertain at $z<4$ due to the small sky coverage and limited blue wavelength coverage of \jwst. We present our ground-based search for LRDs at $2\lesssim z\lesssim4$, combining ultra-deep Hyper Suprime-Cam photometry and various (near-)infrared surveys within a total area of $\sim3.1\,\mathrm{deg^{2}}$. We find that for LRDs with $M_{5500}<-22.5$, their number density declines from $\sim10^{-4.5}\,\mathrm{cMpc^{-3}}$ at $z>4$  to $\sim10^{-5.3}\,\mathrm{cMpc^{-3}}$ at $2.7<z<3.7$ and $\sim10^{-5.7}\,\mathrm{cMpc^{-3}}$ at $1.7<z<2.7$. We also present the Magellan/FIRE spectrum of our first followed-up candidate, \source\ at $z_\mathrm{spec}=2.26$, as a proof of concept for our sample selection. Similar to high-redshift LRDs, the spectrum of \source\ exhibits broad H$\alpha$ emission with $\mathrm{FWHM}\approx2400\,\mathrm{km\,s^{-1}}$ and with nearly symmetric narrow \ha\ absorption. Additionally, \source\ has extremely narrow \oiii\ lines with $\mathrm{FWHM}\approx140\,\mathrm{km\,s^{-1}}$, suggesting the presence of an accreting black hole in a low-mass host galaxy. Limited by the angular resolution of ground-based surveys, we emphasize that spectroscopic follow-ups are required to characterize the contamination fraction of this sample and pin down LRD number density at $z<4$. 
\end{abstract}


\section{Introduction}\label{sec:intro}
One of the most surprising results from early observations with the \textit{James Webb Space Telescope} (\jwst) is the discovery of an abundance of compact red objects at $z>4$ \citep{Kocevski2023, Akins2024, Barro2024, Kocevski2024, Kokorev2024LRD, Labbe2025photLRD}. For their color and morphology, these objects are dubbed ``little red dots" (LRDs, \citealt{Matthee2024LRD}). Their photometric SEDs display a faint blue ultraviolet (UV) continuum along with an extremely red rise in the rest-frame optical, resulting in a characteristic V-shaped SED. 

However, the physical origins of such SED shape and the nature of LRDs remain unclear. Spectroscopic follow-up reveals that $\sim80\%$ of the LRDs show broad \ha\ emission lines with $\mathrm{FWHM}\gtrsim2000\,\mathrm{km\,s^{-1}}$ that are typically seen in type 1 quasars \citep{Kocevski2023, Furtak2024Nature, Greene2024}. Resonating with their compact sizes of $r_\mathrm{e}\lesssim100\,\mathrm{pc}$ \citep{Labbe2025photLRD}, the presence of broad lines sparks the active galactic nucleus (AGN) interpretation of LRDs \citep{Inayoshi2025break, Ji2025blackthunder, Li2025zhengrong}. Yet, the lack of X-rays \citep{Ananna2024xray, Yue2024xray} and mid-infrared (MIR) emission \citep{Williams2024flatMIRI} associated with AGN corona and torus, respectively, challenge the AGN interpretation, although X-ray weak and hot-dust-poor AGN have been observed prior to \jwst\ \citep[e.g.,][]{Lyu2017dustdeficientAGN, Fujimoto2022xrayweakdusty, Ma2024ERQ}. 

Spectroscopic observations of LRDs also frequently reveal that the inflection point of the V-shaped SED consistently occurs at the Balmer limit \citep{Setton2024LRDbreak}, leading to the heavily reddened AGN-galaxy composite models \citep{Killi2024, Labbe2024monster, Wang2024BRD, Ma2025Error404} and even galaxy-only models \citep{Baggen2024} for their SEDs. However, these models often infer uncomfortably large stellar mass and density in the early universe, unusual dust properties, as well as uncertain black hole properties --- \cite{Ma2025Error404}, \cite{deGraaff2025bhstar}, and \cite{Naidu2025BHstar} highlight many of these challenges. Equally important, the lack of far-infrared dust emission also disfavors these models, all of which require significant dust reddening \citep{Labbe2025photLRD,Akins2024, Setton2025ALMA, Xiao2025noCIIdust}. 

Another clue to the nature of LRDs apart from their enigmatic SEDs is their abundance at high redshift. Using a variety of selection methods, different works all estimate a LRD number density of $\sim10^{-4}\,\mathrm{cMpc^{-3}}$ at $z>4$, roughly 100 times more abundant than the UV-selected quasars if extrapolating their number density to similar UV luminosity as the LRDs \citep{Akins2024, Greene2024, Kocevski2024, Kokorev2024LRD}. LRDs also account for a few percent of the high-redshift galaxy population at $-20\lesssim M_{1450}\lesssim-18$ \citep{Greene2024, Kokorev2024LRD}. Yet, individual LRDs and their analogs have only been discovered in small numbers at $z\lesssim4$ \citep{Forrest2024MAGAZ3NE, Juodzbalis2024rossetta, Stepney2024, Wang2024BRD, Lin2025greenpeaLRD}. In addition, most of these searches are not dedicated to LRDs but massive galaxies, red quasars, and emission-line galaxies. \cite{Kocevski2024} carry out a systematic search and identify $\sim300$ LRD candidates at $2\lesssim z\lesssim11$. However, this search only covers $\sim0.18\,\mathrm{deg^2}$ and finds 17 objects below $z=4$, which leaves the number density of LRDs at cosmic noon highly uncertain. 

Therefore, a systematic search for LRDs in a large volume is required to pin down the number density and study their redshift evolution down to cosmic noon. In this paper, we present our search of $2\lesssim z\lesssim4$ LRDs  within the two UltraDeep (UD) fields of the Hyper Suprime-Cam Subaru Strategic Program (HSC--SSP) survey, which contains $\sim3.1\,\mathrm{deg^2}$ of effective sky coverage, nearly 20 times larger than the combined area of the \jwst\ deep fields used in previous studies. We also present an LRD at $z_\mathrm{spec}\approx2.26$, which is photometrically selected and spectroscopically followed up entirely from the ground as a proof of concept for our search. 

The paper is structured as follows. Section~\ref{sec:selection+observation} introduces the parent catalog, sample selection, and observations/reduction of the spectroscopic follow-ups. Section~\ref{sec:number_density} describes our number density estimates of LRD candidates at cosmic noon. Section~\ref{sec:spectrum} presents our spectroscopic follow-up of one LRD candidate. Lastly, we discuss and summarize our results in Section~\ref{sec:summary}. Throughout this work, we assume a cosmology of $\Omega_\mathrm{m,0}=0.3$, $\Omega_\mathrm{\Lambda,0}=0.7$, and $H_0=70\,\mathrm{km\,s^{-1}\,Mpc^{-1}}$. All magnitudes are expressed in AB magnitudes \citep{Oke&Gunn1983} unless specified otherwise.

\section{Sample Selection and Observation}\label{sec:selection+observation}

\subsection{$U$--to--IRAC Catalog}\label{sec:catalog}

We carry out our cosmic noon LRD search and spectroscopic follow-up within all four HSC-SSP Deep/UltraDeep (D/UD) fields. However, specifically for this work regarding the number density estimation, we only focus on the two HSC--SSP UD fields, COSMOS and Subaru Extremely Deep Survey field (SXDS).

These HSC-UD fields have uniform and deep optical imaging with a $5\sigma$ point source detection limit of $i=27.9$ \citep{Aihara2018HSCSSPsurvey}. The optical photometry is supplemented with deep NIR photometry. The COSMOS field overlaps with the Ultra-Deep Survey with the VISTA Telescope (UltraVISTA) with a $5\sigma$ depth of $K_\mathrm{s}=23.7$ \citep{McCracken2012UltraVISTA}. The SXDS field overlaps with the VISTA Deep Extragalactic Observations (VIDEO) survey, which reaches a $5\sigma$ depth of $K_\mathrm{s}=23.5$ \citep{Jarvis2013VIDEO}. Although there are also other NIR surveys covering the two UD fields, for the sake of uniformity in the NIR bands, we restrict our survey area to the overlap between the UltraVISTA/VIDEO footprints and the HSC--SSP UD pointings --- this totals a sky coverage of $3.11\,\mathrm{deg^2}$, which is $\sim17$ times larger than the area searched by \cite{Kocevski2024}. In addition, $U$--band imaging data is provided by the Canada--France--Hawaii Telescope (CFHT) Large Area $U$--band Deep Survey (CLAUDS, \citealt{Sawicki2019CLAUDS}). \textit{Spitzer}/IRAC imaging data at 3.6\,$\mathrm{\mu}m$ and 4.5\,$\mathrm{\mu m}$ are also available in the UD fields from multiple surveys including the Spitzer COSMOS survey (S-COSMOS, \citealt{Sanders2007scosmos}), the Spitzer Extragalactic Representative Volume Survey (SERVS, \citealt{Mauduit2012servs}), the Spitzer DeepDrill survey \citep{Lacy2021spitzerdeepdrilling}, and the Spitzer Large Area Survey with Hyper Suprime-Cam (SPLASH, \citealt{Mehta2018splash}). For the IRAC coverage in COSMOS, we make use of the master IRAC mosaics from the Spitzer Coverage of the HSC-Deep with IRAC for Z-studies (SHIRAZ, \citealt{Annunziatella2023shiraz}) which co-adds new and archival observations. For a complete list of the {\sl Spitzer} observations included in the COSMOS master mosaic, see Table\,2 in \citet{Annunziatella2023shiraz}. For the IRAC coverage in SXDS, we make use of the IRAC mosaic combining all available data from \citet{Lacy2021spitzerdeepdrilling}.

The photometric catalog is defined following the HSC pipeline approach in the HSC fields \citep{Bosch2018HSCpipe, Bosch2019lsstpipe} with the addition $U$ and $JHK/K_\mathrm{s}$ imaging and the needed modification for the pipeline to account for these non-HSC data \citep{Desprez2023combineHSC}. This catalog is called the ``$U$-to-$K/K_\mathrm{s}$ catalog". Peaks are identified in images of $U$--band through $K/K_s$--band simultaneously. The first band in the order of $i$--$r$--$z$--$y$--$g$--$J$--$H$--$K/K_\mathrm{s}$ where a source shows a $7\sigma$ peak is designated as the detection band. Forced photometry is then performed using the same profiles and metrics as given in the third data release of HSC-SSP \citep{Aihara2022HSCDR3}. 

The \textit{Spitzer}/IRAC photometry was performed using an established source fitting algorithm developed to extract photometry for heavily blended/confused images. Briefly, the code utilizes the source segmentation map and image in the higher resolution $K$/$K_\mathrm{s}$--, $J$-- or $z$--bands (prioritizing $K$/$K_\mathrm{s}$ if available, then defaulting to $J$, then $z$) as a prior, to create a model at the resolution of \textit{Spitzer}/IRAC, which is used to subtract nearby/blended sources before forced photometry is performed, using a $D = 3\arcsec$ aperture.  A more in-depth description of the process is provided in Section A.4 of \citet{Marchesini2009}, with the exception that we correct the $D = 3\arcsec$ fluxes to total flux by assuming a median curve of growth, derived from point-like sources in the mosaics, for each field.

The final compiled photometric catalog is referred to as the ``$U$--to--IRAC catalog" in the subsequent text of this work. Due to the nature of forced photometry and the fact that the detection band is usually one of the HSC bands, the effective depth of the NIR photometry could be $\sim2\,\mathrm{mag}$ deeper than the published nominal survey depth fo blind point source detections \citep{McCracken2012UltraVISTA, Jarvis2013VIDEO}. However, the source completeness does drop below those nominal blind detection limits. 

\subsection{Sample Selection}\label{sec:selection}

\begin{figure*}[!ht]
    \centering
    \includegraphics[width=\textwidth]{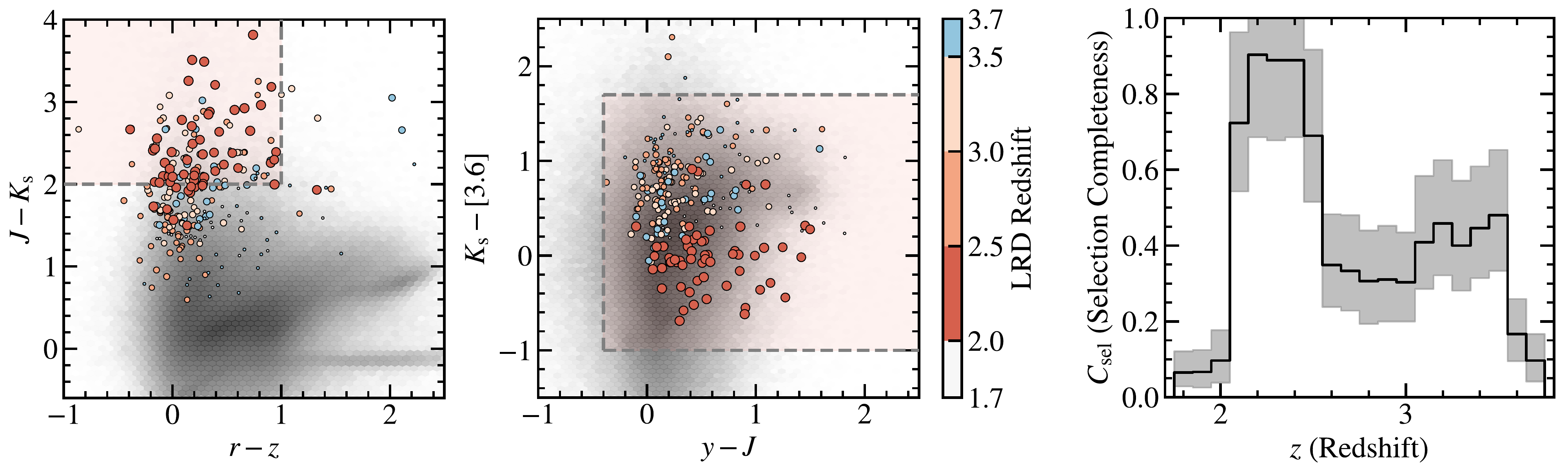}
    \caption{\textbf{Left:} The synthetic photometric colors of the LRD templates \citep{Setton2024LRDbreak} shifted to various redshift are shown in the UV-optical color space. The sizes represent the selection completeness $C_\mathrm{sel}$ at the redshift. The selection boxes are shaded red. All sources in the $U$--to--IRAC catalog are plotted as hexagons in the background. \textbf{Middle:} Same as the left panel, but instead in the optical-NIR color space. \textbf{Right:} The completeness of the proposed color selection is shown as a function of redshift. The gray shaded regions are $1\sigma$ Poisson uncertainties. }
    \label{fig:color_selection}
\end{figure*}

Selections for LRDs at $5<z<8$ with \jwst\ are all designed to target the characteristic V-shaped UV-optical SED. At $2\lesssim z\lesssim4$, the rest-frame UV shifts into the observed optical bands, which is covered by the five-band HSC--$grizy$ photometry. The rest-frame optical shifts into the observed NIR, where we have coverage in the NIR bands. The rest-frame NIR can also be covered by the \textit{Spitzer}/IRAC bands. 

Due to the lack of uniform $H$--band coverage in the full HSC--Deep/UltraDeep fields, we carry out a simple color-color selection approach \citep[e.g.,][]{Akins2024, Greene2024, Kokorev2024LRD, Labbe2025photLRD} instead of a photometric slope approach like in \cite{Kocevski2024}. We propose the following criteria\footnote{In all the selection criteria, $K_\mathrm{s}$ can be replaced by $K$. We use $K_\mathrm{s}$ because NIR imaging data in the UltraDeep fields are from UltraVISTA and VIDEO, which we use to estimate number density.} to select LRDs at cosmic noon: 
\begin{itemize}
    \item signal-to-noise ratio is greater than 3 in all bands from $r$ to $K_\mathrm{s}$;
    \item the object is consistent with a point source (i.e., $m_\mathrm{PSF}-m_\mathrm{CMODEL}<0.0164$; \citealt{Bosch2018HSCpipe}) in the $i$, $J$, $H$, and $K_\mathrm{s}$ bands;
    \item $J-K_\mathrm{s}>2$ for the red rest-frame optical emission;
    \item $r-z<1$ for the blue rest-frame UV emission. 
\end{itemize}
The last two color cuts mimic the $\mathrm{F277W-F444W}$ and $\mathrm{F115W-F200W}$ color cuts proposed by \cite{Greene2024} for selecting $z>5$ LRDs with \jwst/NIRCam filters. Additionally, we also require two supplementary color cuts:
\begin{itemize}
    \item $y-J > -0.4$,
    \item $-0.8 <K_\mathrm{s}-[3.6] < 1.8$ if \textit{Spitzer}/IRAC photometry is available.
\end{itemize}
The first of these two color cuts is to remove poor photometry with blue optical-NIR color that is inconsistent with that of LRDs. Such colors mostly arise in sources at $y\gtrsim25.8$, where the forced photometry goes deeper than the $3\sigma$ depth of blind source detection. The $K_\mathrm{s}-[3.6]$ color is not explicitly included in the high-$z$ selection mostly due to lack of deep \textit{JWST}/MIRI coverage in these fields at the time that LRDs were discovered. Given the very limited number of spectroscopically confirmed LRDs with MIRI photometry \citep{Labbe2024monster, Wang2024BRD, Setton2025ALMA}, the variation of rest-frame optical-NIR color is not as well constrained as the rest-frame UV-optical distribution. Therefore, the color range we assume is relatively forgiving. Extremely red quasars and hot dust-obscured galaxies at cosmic noon, whose optical-NIR SEDs rise steeply \citep[e.g.,][]{Tsai2015HotDOG, Hamann2017ERQ}, would be excluded. Objects with unphysically blue optical-NIR color due to poor photometry would also be excluded. 

To test our low-redshift color cuts, we adopt the sample of 28 little red dots spectroscopically identified by \cite{Setton2024LRDbreak}. Although the sample is initially selected for extremely red \ha\ emitters, spectroscopically identified LRDs from various works \citep[e.g.,][]{Greene2024, Juodzbalis2024rossetta, Kocevski2024, Wang2024BRD} do fall in that selection box. To further include LRDs with strong Balmer breaks, we also include Abell\,2744-QSO1 at $z=7$ \citep{Furtak2024Nature}. These templates are shifted within $1.7<z<3.7$ with a redshift increment of $\Delta z=0.1$, and the synthetic $J-K_\mathrm{s}$, $r-z$, $y-J$, and $K_\mathrm{s}-[3.6]$ colors are computed for each object at each redshift. We show in Figure~\ref{fig:color_selection} the positions of these synthetic photometric colors relative to our proposed color selection boxes and the full sample from the $U$--to--IRAC catalogs. Qualitatively, our rest-frame UV-optical color cuts are indeed able to select LRDs at various redshifts, and the rest-frame optical-NIR color cut is sufficiently forgiving to account for LRDs at nearly all redshifts. Quantitatively, we also estimate the selection completeness $C_\mathrm{sel}$ as the fraction of LRDs at each redshift that meet the selection criteria, out of a total of 29. We also show the $C_\mathrm{sel}$ as a function of redshift in Figure~\ref{fig:color_selection}. As evident from Figure~\ref{fig:color_selection}, our color selection is most effective in selecting LRD candidates at $z\approx2.0$--2.5. However, this selection completeness drops to approximately half at $z\sim3$ as $J$--band shifts to blueward of the Balmer break and \ha\ shifts out of $K_\mathrm{s}$--band, resulting in a bluer $J-K_\mathrm{s}$ color. The completeness also plummets at $z\lesssim2$ as \oiii\ shifts into $J$--band while $K_\mathrm{s}$--band approaches NIR wavelength, flattening the rest-frame optical SED. 

In total, we identified 342 LRD candidates in the four HSC-Deep fields, 182 of which located in the HSC-UltraDeep footprints. We provide the list of the UltraDeep sample in Table~\ref{tab:sample} of Appendix~\ref{app:sample}. 

\subsection{Spectroscopic Follow-up}
\begin{figure}[!ht]
    \centering
    \includegraphics[width=\columnwidth]{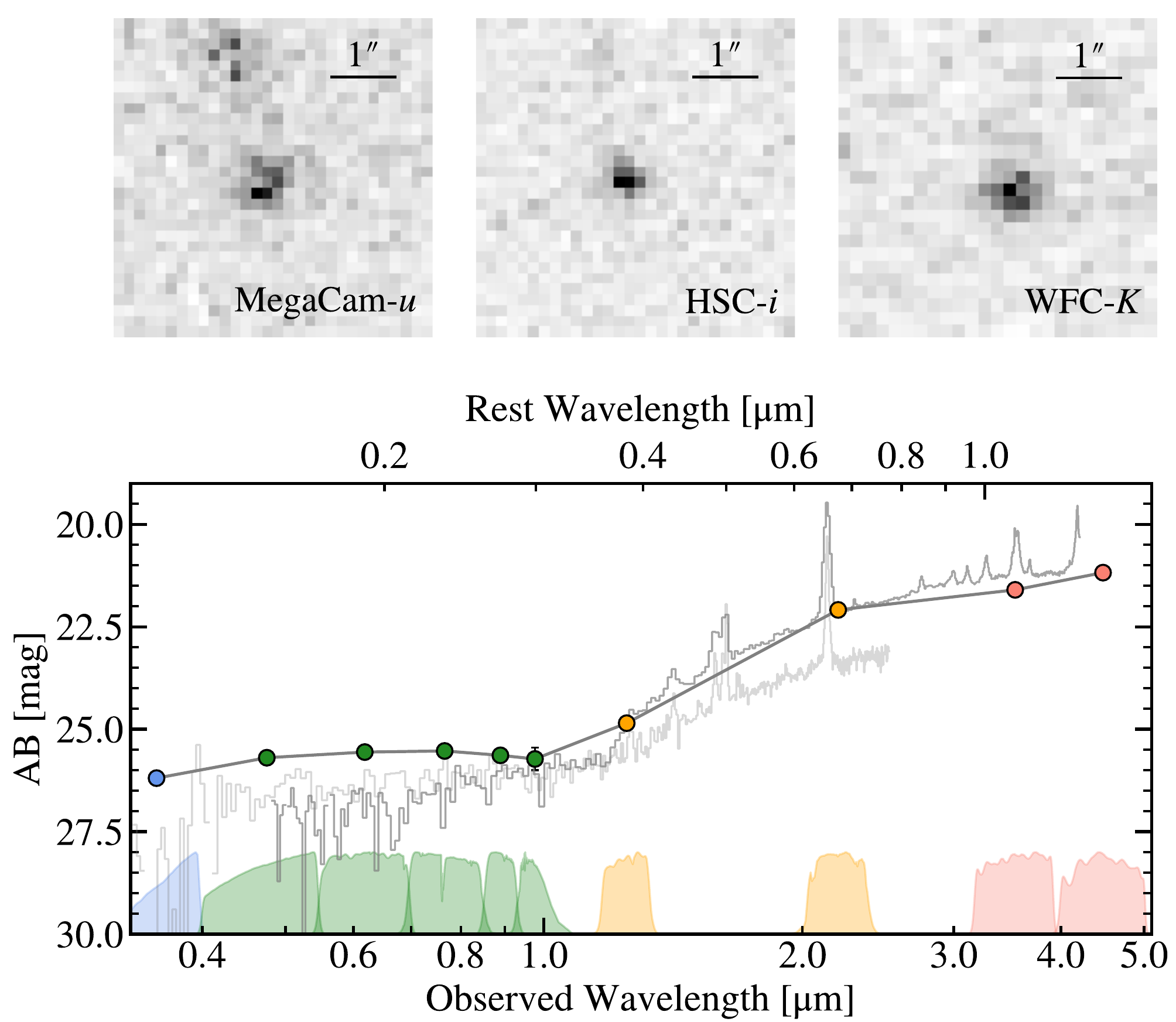}
    \caption{\textbf{Upper:} The $5\arcsec$$\times$$5\arcsec$ cutout images of \source\ in the $u$--, $i$--, and $K$--bands. \textbf{Lower:} The photometric SED of \source{} from the CFHT-$U$ band to \textit{Spitzer}/IRAC 4.5\,$\mathrm{\mu m}$ (\citealt{Timlin2016SpIES}) is shown against the rescaled \jwst/NIRSpec spectrum of UNCOVER-4286 (light gray) and RUBIES-BLAGN-1 (dark gray), two little red dots at $z\approx5.8$ and $z\approx3.1$, respectively \citep{Greene2024, Wang2024BRD}. We also plot the transmission curves of the photometric filters corresponding to each photometric point.}
    \label{fig:phot_SED}
\end{figure}

In Section~\ref{sec:spectrum}, we present in detail our first spectroscopic follow-up observation of a cosmic noon LRD candidate, \source. We briefly describe the observation and data reduction here in this section. Although we restrict ourselves to the HSC-UltraDeep fields for number density estimation, our spectroscopic follow-up campaign extends to all four HSC-Deep fields (e-COSMOS, XMM-LSS, ELAIS-N1, DEEP2-F3). A $U$-to-IRAC catalog is constructed for each of the four HSC-Deep fields following the same procedure as described in Section~\ref{sec:catalog} (also see \citealt{Aihara2022HSCDR3} for detailed description of the multiwavelength imaging in all four HSC-Deep fields). In the DEEP2-F3 field, where \source\ is selected from, the $J$-- and $K$--band imaging are from the Deep UKIRT NEar-Infrared Steward Survey (DUNES$^2$; \citealt{Aihara2022HSCDR3}, Egami et al., in prep.). LRD selection is done in the same way as described in Section~\ref{sec:selection}.

We choose \source\ as the first candidate to be followed up for two reasons. Firstly, it is the reddest target among all the candidates in the DEEP2-F3 fields that is bright enough ($K\approx22.1$\,mag) to make an \ha\ detection more likely and feasible with limited on-source time. Secondly, \source\ has a large optical-UV color contrast (see Figure~\ref{fig:sample}). We show the UV-to-NIR SED of \source\ in Figure~\ref{fig:phot_SED}, which also visualizes our selection criteria. The follow-up observation was done with the Folded-port InfraRed Echellette (FIRE) on the Magellan Baade Telescope on October 18--19, 2024. We conducted an ABBA nodding pattern for NIR difference imaging. Individual exposures are 1200 seconds each, and all eight science frames totals 160 minutes of integration. We used a slit width of $0.\arcsec75$.

We reduce the FIRE spectrum with \texttt{PypeIt}, a python-based reduction pipeline \citep{Prochaska2020JOSSpypeit, Prochaska2020zenodoPyPeit}. For each frame, we calibrate the wavelength solution using the prominent OH emission lines from the sky. No continuum emission is visually obvious in the AB-subtracted 2D spectra for the pipeline to fully automate trace detection. We therefore manually identify emission lines and force the pipeline to extract at specific locations of each echelle order. The extraction uses a generalized \cite{Horne1986} method with $\mathrm{FWHM=5\,pixels}$. Lastly, we perform flux calibration and telluric corrections using the spectrum of the telluric standard star obtained on the night of observation. 

Overall, we model emission lines in the reduced spectrum with Gaussian profiles. We allow both permitted and forbidden lines to have a component with $\mathrm{FWHM<2000\,\mathrm{km\,s^{-1}}}$ \citep{Zakamska2003} to represent narrow line region (NLR) emission. For the permitted lines, we also allow a broad component with $\mathrm{FWHM>2000\,\mathrm{km\,s^{-1}}}$ to represent emission from the broad line regions (BLR). We also allow one kinematically offset \ha{} absorption feature in the model --- those spectral features are observed in many high-$z$ LRDs \citep[e.g.,][]{Matthee2024LRD, Ji2025blackthunder}. Except for the absorption features, all other components are fixed at the same systemic velocity. We also fit for a constant offset for each emission line to account for imperfect sky subtraction in the reduction process and/or weak continuum emission.

\section{Number Density at Cosmic Noon}\label{sec:number_density}

\subsection{LRD Candidates and Photometric Redshifts}

Using the selection criteria outlined in Section~\ref{sec:selection}, we identified 182 LRD candidates within 3.1\,square degrees of the two HSC-UD fields, out of a total of 342 LRD candidates across all four HSC-Deep fields. We provide a full list of the sample in Table~\ref{tab:sample} of Appendix~\ref{app:sample}. Now, we investigate the number density of LRDs at $z<4$ and compare with their high-redshift counterparts selected by \jwst. We note that we restrict the following number density estimation only within the UD fields, where the NIR photometry is most uniform and the optical photometry is the deepest, which is crucial to detect the faint rest-frame UV.

We derive the photometric redshift of the LRD candidates using the Python version of Easy and Accurate $z_\mathrm{phot}$ from Yale (\texttt{EAZY}, \citealt{Brammer2008eazy}). We supplement the default \texttt{tweak\_fsps\_QSF\_12\_v3} template set with the UV-to-NIR SED of two LRDs with complete UV-to-NIR modeling, RUBIES-BLAGN-1 \citep{Wang2024BRD} and UNCOVER-45924 \citep{Labbe2024monster} to account for the high-equivalent-width \ha\ emission lines and the red continuum of LRDs not accounted for by the default set. Similar to \cite{Kocevski2024} and \cite{Kokorev2024LRD}, we assume a flat redshift prior between 0 and 5. Our $z_\mathrm{phot}$ estimation is not sensitive to the choice of the prior upper bounds. Only 17 objects fall out of the $1.7<z<3.7$ window that we focus on, yielding 165 objects for number density estimation. 

\subsection{Luminosity Function}\label{sec:LF}
\begin{figure*}
    \centering 
    \includegraphics[width=\textwidth]{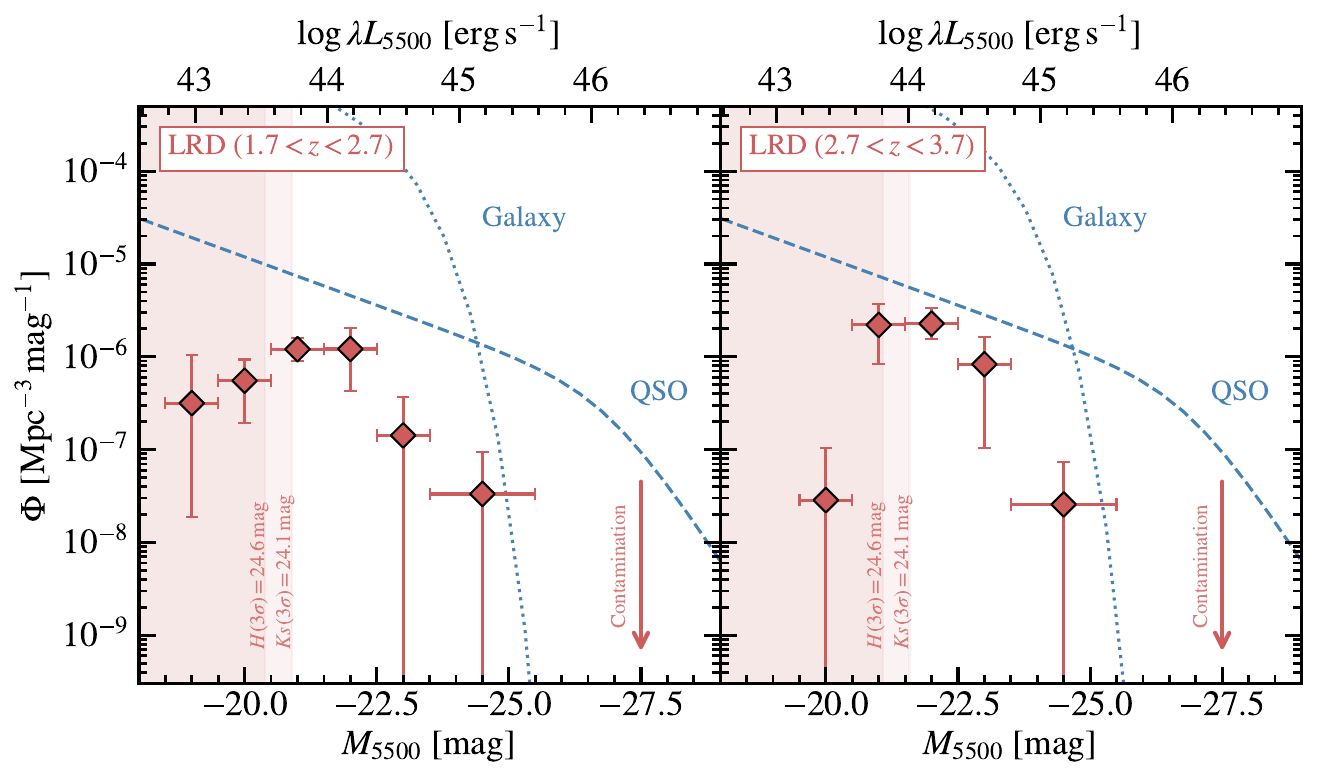}
    \caption{The optical luminosity functions of LRD candidates at $1.7<z<3.7$ are shown in red diamonds. The red shaded regions are the luminosities corresponding to the $H$ and $K_\mathrm{s}$ $3\sigma$ detection limits for blind point source detection at $z=2.2$ and $z=3.2$. Although we detect sources below those limit due to the forced photometry nature of the $U$--to--IRAC catalogs (see Section~\ref{sec:catalog}), the selection is likely rather incomplete at those luminosities. The observed quasar luminosity (QLF) functions at $2.2<z<3.5$ \citep{Ross2013QLF_SDSS} is shown as the blue dashed curve. The galaxy LFs at $2.1<z<2.7$ and $2.7<z<3.3$ \citep{Marchesini2012galaxyLF} are shown as the blue dotted curve.}
    \label{fig:lf}
\end{figure*}

In this work, instead of computing the UV luminosity function (UVLF) as is commonly done in previous LF studies \citep{Greene2024, Kocevski2024, Kokorev2024LRD}, we choose to compute the optical luminosity function of LRDs, which is characterized by the absolute magnitude at rest-frame 5500\,\AA, $M_{5500}$. The UVLF choice is mostly for more direct comparison with luminous quasars at comparable redshifts, which were most easily observed in the rest-frame UV wavelengths prior to \textit{JWST} \citep[e.g.,][]{Ross2013QLF_SDSS, Akiyama2018z4QLF, Matsuoka2023z7QLF}. However, the UVLF of LRDs is subject to their selection function. Since many LRD SEDs are flat (in $F_\nu$ units) in the UV, their UV luminosity is largely determined by their rest-frame optical luminosity and their UV-optical color --- the latter is arbitrarily bounded by the selection criteria. In the context of our selection, these quantities would be $M_{K_\mathrm{s}}$ and the $J-K_\mathrm{s}$ color. Additionally, unlike the type 1 quasars, the bolometric correction of LRDs remains uncertain \citep{Kocevski2023, Matthee2024LRD}. Therefore, we consider that a rest-frame optical LF would be a sensible choice less dependent on the selection criteria. 

To compute the optical luminosity function, we employ a modified version of the $1/V_\mathrm{max}$ method \citep{Schmidt1968}. The $1/V_\mathrm{max}$ method does not assume any intrinsic shape for the LF, which is suitable for LRDs as their luminosity distribution remains uncertain. The number density $\Phi(M)\Delta M$ of the LRD candidates in each absolute magnitude bin $\Delta M$ is estimated as
\begin{equation}\label{eq:1/vmax}
    \Phi(M)\Delta M = \sum_i\left[\frac{P_{\mathrm{color},i}}{C_\mathrm{sel}(z_i)}\cdot\frac{1}{V_\mathrm{max}(A, z_\mathrm{min}, z_{\mathrm{max},i})}\right]\,\,,
\end{equation}
where $i$ indexes each object within the bin, $P_\mathrm{color}$ is the probability that an LRD candidate falls into the color box given its photometric uncertainties, $C_\mathrm{sel}(z)$ is the selection completeness at redshift $z$ presented in Figure~\ref{fig:color_selection}, and $V_\mathrm{max}$ is the maximum comoving volume within the survey area $A$ that a source can be detected. The extra 1/$C_\mathrm{sel}$ factor (ranging from $\sim1$--2.5) from the standard $1/V_\mathrm{max}$ accounts for the under-counting by our selection criterion with respect to the \cite{Setton2024LRDbreak} reference sample. In detail, $V_\mathrm{max}$ for each source explicitly depends on the lower bound of the concerned redshift range, $z_\mathrm{min}=1.7$, and   
\begin{equation}\label{eq:zmax}
    z_{\mathrm{max},i} = \mathrm{min}(z_{\mathrm{lim},i}, z=3.7), 
\end{equation}
where $z_\mathrm{lim}$ is the largest redshift to detect a source in a photometric band $X$ given the detection limit $m_{X,\mathrm{lim}}$ and the absolute magnitude $M_X(z_i)$ derived from the source redshift $z_i$. 

We compute the optical luminosity function of the LRD candidates in a Monte-Carlo fashion with 1000 realizations. For each realization, we randomly draw the photometric redshifts from each LRD's $p(z)$ distribution. Based on the redshift, we draw the $C_\mathrm{sel}$ values according to their uncertainties. We also compute $M_{5500}$ by interpolating between the NIR photometric points, each drawn according to their photometric errors. All optical luminosities are then binned, and number densities and corresponding uncertainties arising from redshift and selection completeness are computed based on Equations~\ref{eq:1/vmax} and \ref{eq:zmax}. The estimated uncertainties are then convolved with the Poisson errors. We also repeat the calculation for the two UD fields individually and estimate the error due to cosmic variance as half of the difference between the two fields. The cosmic variance is then added in quadrature with the statistical error to obtain final errors for the optical luminosity function, which we also report in Table~\ref{tab:luminosity_function}. 

\begin{deluxetable}{ccc}[!ht]\label{tab:luminosity_function}
\tablecaption{Optical luminosity function of the LRD candidates at intermediate redshift}
\tablewidth{0pt}
\tablehead{
\colhead{$M_{5500}$} & \colhead{$\log\Phi$ ($1.7<z<2.7$)} & \colhead{$\log\Phi$ ($2.7<z<3.7$)} \\
\colhead{[mag]} & \colhead{[$\mathrm{cMpc^{-3}\,mag^{-1}}$]} & \colhead{[$\mathrm{cMpc^{-3}\,mag^{-1}}$]}
}
\startdata
$-19.0$ & $-6.50_{-1.23}^{+0.52}$ & \nodata \\
$-20.0$ & $-6.26_{-0.45}^{+0.23}$ & $-7.54_{-\infty}^{+0.56}$ \\
$-21.0$ & $-5.92_{-0.13}^{+0.12}$ & $-5.66_{-0.43}^{+0.22}$ \\
$-22.0$ & $-5.92_{-0.46}^{+0.23}$ & $-5.64_{-0.16}^{+0.17}$ \\
$-23.0$ & $-6.85_{-\infty}^{+0.41}$ & $-6.08_{-0.91}^{+0.30}$ \\
$-24.5$ & $-7.48_{-\infty}^{+0.45}$ & $-7.59_{-\infty}^{+0.46}$ \\
\enddata
\end{deluxetable}

We show the optical luminosity function of the LRD candidates at cosmic noon in Figure~\ref{fig:lf} and Table~\ref{tab:luminosity_function}. We also put our LRD optical luminosity function in the context of galaxy and quasar populations by comparing with their respective LFs at similar redshift. However, the QLF at cosmic noon is often represented as the UVLF (e.g., $M_{i,z=2}$ in \citealt{Ross2013QLF_SDSS}), which necessitates a conversion between optical to UV luminosity for these high-redshift quasars. By conducting HSC--$i$ synthetic photometry on the $z=2$ type 1 quasar SED template provided by \cite{Temple2021}, we derive the UVLF-to-optical luminosity function conversion as $M_{5500} = M_{i,z=2}+ 0.08$. 
As shown in Figure~\ref{fig:zevolution}, the number density of the V-shaped LRD candidates are lower than that of the quasars, and they are less abundant than galaxies at comparable optical luminosity by nearly a factor of $\sim1000$. In contrast, at $z\sim4$--8, LRDs are $\sim100$ times more abundant than the UV-selected quasars and accounts for a few percent of the galaxy populations at that cosmic time \citep{Akins2024, Greene2024, Kocevski2024, Kokorev2024LRD}. 

With the $3.1\,\mathrm{deg^2}$ sky coverage, our survey area is more than an order of magnitude larger than that of the search within \jwst\ fields ($0.18\,\mathrm{deg^2}$; \citealt{Kocevski2024}). The larger volume thus allows us to more easily probe the bright end of the LF ($M_{5500}\lesssim-23$) not easily characterized by \jwst\ (see \citealt{Kocevski2024} and \citealt{Kokorev2024LRD}) for the intrinsic rarity of these most luminous candidates. On the bright end, the optical luminosity function of our LRD candidates may tentatively drop at approximately $M_{5500}\approx-25$, whereas the optical luminosity function of UV-selected quasars turns over at $M_{5500}\approx-27$. 

Nevertheless, we reiterate that the LF we present should be regarded as an upper limit. Dusty star-forming galaxies such as ULIRGs could also enter the color selection boxes of LRDs (see Figure~\ref{fig:contamination} in Appendix~\ref{app:contamination}) at similar redshift. Their number density is not well constrained. The HSC--$i$ band has a point spread function of $0\arcsec.61$ \citep{Aihara2022HSCDR3}, the smallest of all bands in our catalog. Nonetheless, this size still translates into $r_\mathrm{e}\approx4$--5\,kpc at $1.7<z<3.7$, which a compact dusty star-former could fit within \citep[e.g.,][]{Cheng2023SMG, Huang2023z2ULIRG}. Our ongoing spectroscopic follow-up campaign that searches for broad emission lines, which are uniquely present in LRDs, could characterize the contamination fraction in the future. 

\subsection{Redshift Evolution}
\begin{figure}[!ht]
    \centering 
    \includegraphics[width=\linewidth]{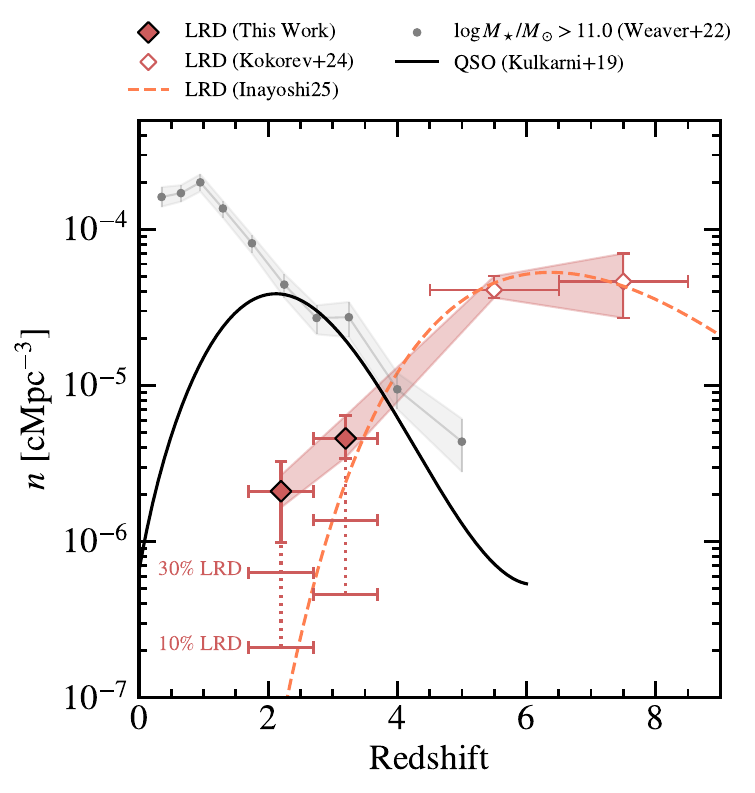}
    \caption{The number density of LRDs ($M_{5500}<-20.5$) is shown as a function of redshift. The solid red diamonds represent the values estimated in this work, and the empty red diamonds are derived from \cite{Kokorev2024LRD}. We show the number density evolution of quasars as the solid curve \citep{Kulkarni2019QLFzevo} by integrating the quasar UVLF at $M_{1450}<-20.1$, which corresponds to $M_{5500}<-20.5$ according to the \cite{Temple2021} template. The number density evolution of galaxies is shown as gray circles \citep{Weaver2023MF}. We also show the LRD number density evolution suggested by \cite{Inayoshi2025lognormal}. The horizontal bars represent number density estimation of different contamination fraction. }
    \label{fig:zevolution}
\end{figure}

By integrating the optical luminosity function, we are also able to estimate the total number density of the selected LRD candidates ($M_{5500}<-20.5$) at cosmic noon and cosmic dawn. We estimate the number density of LRDs to be $2.1_{-1.1}^{+1.1}\times10^{-6}\,\mathrm{cMpc^{-3}}$ at $1.7<z<2.7$ and $4.6_{-1.2}^{+1.8}\times10^{-6}\,\mathrm{cMpc^{-3}}$ at $2.7<z<3.7$. We present the redshift evolution of LRD number density in Figure~\ref{fig:zevolution}. We find that the number density of LRDs drops nearly by a factor of $\sim10$ from $z>4$ to $z<4$. This is qualitatively consistent with the finding by \cite{Kocevski2024} that the total number of LRDs within \jwst\ deep fields decreases significantly as the search approaches cosmic noon. Once again, we reiterate that this number density should be viewed as an upper limit for the contamination argument discussed in Section~\ref{sec:LF}. 

The behavior of the redshift evolution of LRD number density is also drastically different from those of quasars and massive galaxies as we show in Figure~\ref{fig:zevolution}. While LRDs become increasingly rare from $z\sim6$ to $z\sim2$, quasar number density monotonically increases from $z\sim6$ and peaks at $z\sim2$ before dropping off in the modern-day universe \citep{Kulkarni2019QLFzevo, Shen2020QLF}, evolving in the opposite way as that of LRDs. Galaxies with stellar masses above $10^{11}\,M_\odot$ also show the opposite evolution as our LRDs. 
Our number density at $z\sim3$ is consistent with that of \cite{Inayoshi2025lognormal}, who assumes that the occurrence rate of LRDs across $2\lesssim z \lesssim 11$ found by \cite{Kocevski2024} follows a log-normal distribution. However, our estimated number density of $z\sim2$ LRD candidates is higher than the prediction of \cite{Inayoshi2025lognormal} by nearly an order of magnitude. This discrepancy may be alleviated due to the unknown contamination fraction that can only be characterized with spectroscopic follow-ups.

\section{A Case Study Validation}\label{sec:spectrum}

\begin{deluxetable}{lll}
\tablecaption{Photometric and Kinematic Measurements of \source}
\tablewidth{0pt}
\tablehead{
\colhead{Quantity} & \colhead{Unit} & \colhead{Measurements}
}
\startdata
RA & deg & $351.8995$ \\
Dec & deg & $-0.2282$ \\
$z_\mathrm{spec}$ & --- & $2.2605$ \\
\hline
CFHT/MegaCam--$u$ & mag & $26.19\pm0.04$ \\
Subaru/HSC--$g$ & mag & $25.70\pm0.03$ \\
Subaru/HSC--$r$ & mag & $25.56\pm0.04$ \\
Subaru/HSC--$i$ & mag & $25.53\pm0.06$ \\
Subaru/HSC--$z$ & mag & $25.64\pm0.11$ \\
Subaru/HSC--$y$ & mag & $25.73\pm0.28$ \\
UKIRT/WFC--$J$ & mag & $24.85\pm0.16$ \\
UKIRT/WFC--$K$ & mag & $22.09\pm0.02$ \\
\textit{Spitzer}/IRAC\,[3.6] & mag & $21.60\pm0.04$ \\
\textit{Spitzer}/IRAC\,[4.5] & mag & $21.18\pm0.04$ \\
\hline
$\log L_\text{\ha,BL}$ & $\mathrm{erg\,s^{-1}}$ & $43.25_{-0.01}^{+0.02}$ \\
$\log L_\text{\ha,NL}$ & $\mathrm{erg\,s^{-1}}$ & $42.43_{-0.03}^{+0.03}$ \\
$\log L_\text{\oiii}$ & $\mathrm{erg\,s^{-1}}$ & $42.63_{-0.01}^{+0.01}$ \\
$\mathrm{FWHM_{BL}}$ & $\mathrm{km\,s^{-1}}$ & $2447_{-117}^{+105}$ \\
$\mathrm{FWHM_{NL}}$ & $\mathrm{km\,s^{-1}}$ & $144_{-3}^{+3}$ \\
$\mathrm{FWHM_{abs}}$ & $\mathrm{km\,s^{-1}}$ & $478_{-79}^{+91}$ \\ 
$\Delta v_\mathrm{abs}$ & $\mathrm{km\,s^{-1}}$ & $-52_{-26}^{+21}$ \\ 
$\mathrm{REW_{abs}}$ & $\mathrm{\AA}$ & $6.4_{-1.0}^{+1.1}$ \\
\enddata
\tablecomments{The\ha\ luminosity presented in the table does not take into account the narrow absorption features.}
\label{tab:line_fits}
\end{deluxetable}

\begin{figure*}
    \centering
    \includegraphics[width=\textwidth]{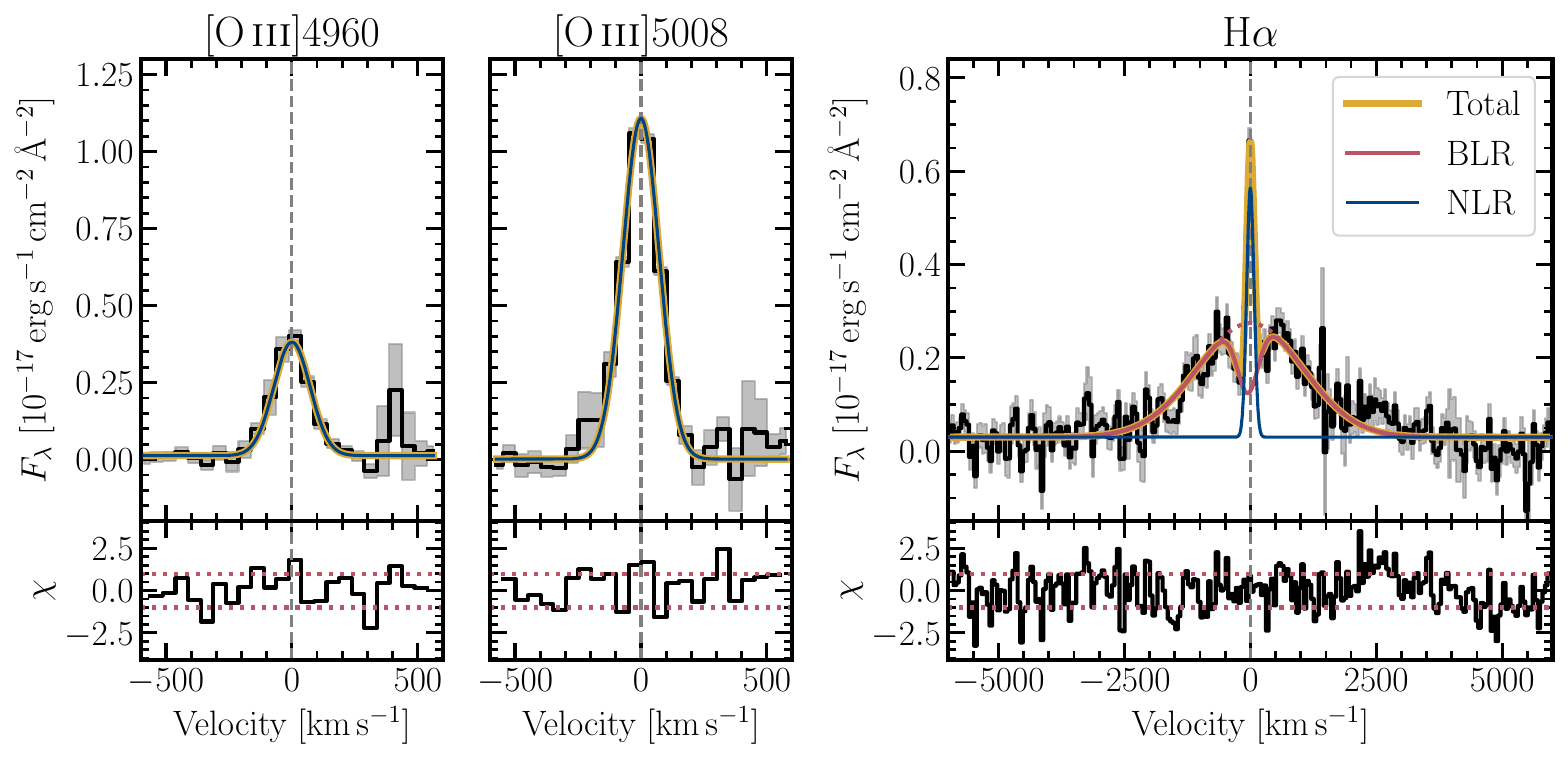}
    \caption{The emission lines detected in the FIRE spectrum of DEEP23-z2LRD1 (one of HSC-Deep fields, not UltraDeep) is plotted in velocity space. The yellow curve represents the total model. The blue curve represents the NLR emission. The solid and dashed red curves represents the BLR emission with and without the \ha{} absorptions features included, respectively. The gray shaded area is the $1\sigma$ measurement uncertainty associated with the data.}
    \label{fig:emission_lines}
\end{figure*}
Having discussed the $2\lesssim z\lesssim4$ LRD candidates as a population, we now turn to a case study of a broad-line LRD at $z_\mathrm{spec}=2.2605$, discovered and spectroscopically confirmed entirely from the ground. \source\ was observed as the very first object in our ongoing spectroscopic follow-up campaign of cosmic noon LRD candidates to confirm their redshifts and detect broad emission lines. We present the $U$--to--IRAC photometry and the best-fit kinematic measurements of \source\ in Table~\ref{tab:line_fits}. 

As easily seen from the FIRE spectrum in Figure~\ref{fig:emission_lines}, we confirm the broad-line nature of \source\ from our best-fit model. We measure the broad-line component of \ha\ emission line to have $\mathrm{FWHM}=2447_{-117}^{+105}\,\mathrm{km\,s^{-1}}$. On top of the broad line, we also detect a nearly symmetric narrow \ha\ absorption --- commonly seen in high-resolution spectra of high-redshift LRDs \citep{Matthee2024LRD, Lin2024xiaojing, Wang2024BRD} --- with a rest-frame equivalent width (REW) of $6.4_{-1.0}^{+1.1}\,\mathrm{\AA}$. We do not detect \hb\ emission. It is most likely that an OH emission line from the sky contaminates an intrinsically weak \hb\ line. However, this could be due to a large Balmer decrement due to heavy reddening or collisional de-excitation of \hb\ in the broad line region \citep{Korista&Goad2004BLRlines}. We also do not detect the \nii$\lambda\lambda6550,6585$ doublet, which may be due to the limited SNR of the spectrum. Based on the \ha\ line width and luminosity, we invoke the standard scaling relation from \cite{Greene&Ho2005} and estimate a black hole mass\footnote{We do not correct for dust reddening when estimating $M_\mathrm{BH}$ given the non-detection of FIR dust emission in bright high-redshift LRDs \citep{Setton2025ALMA}.} of $M_\mathrm{BH}=6.18_{-1.22}^{+1.25}\times10^7\,\mathrm{M_\odot}$, similar to those of the high-$z$ LRDs \citep{Greene2024}. Of course, as noted by many works, the scaling relation is only calibrated in the local universe, and it likely only provides an upper limit for the black hole masses in LRDs given their compactness and potential high Eddington-ratios leading to smaller BLR \citep{Lupi2024BLRsize, Maiolino2025xray}. If taking the $M_\mathrm{BH}$ at face value, we also estimate an $L/L_\mathrm{Edd}\approx0.3$ based on the bolometric correction from \cite{Stern&Laor2012bolcorr}. 

Leveraging the high spectral resolution ($R\approx6000$) of Magellan/FIRE, we are also able to resolve the narrow \oiii$\lambda\lambda4960,5008$ doublets and the narrow \ha. We measure an extremely narrow line width of $\mathrm{FWHM}=144\pm3\,\mathrm{km\,s^{-1}}$ for the narrow lines, accounting the instrumental resolution from nearby OH emission lines. We can also estimate a dynamical mass of the host galaxy with
\begin{equation}
    M_\mathrm{dyn} = \frac{Kr_\mathrm{e}\sigma_v^2}{G}\,\,,
\end{equation}
where $\sigma_v$ is the velocity dispersion, $r_\mathrm{e}$ is the effective radius, $G$ is the gravitational constant, and $K$ is a constant that depends on the brightness profiles (typically between 3--5; see \citealt{Cappellari2006}, \citealt{vanderWel2006Mdyn}, \citealt{Taylor2010Mdynlocal}), which we take to be 5. Since \source\ is unresolved, large uncertainties in the dynamical mass are associated with its size. If taking the $i$--band PSF size as an upper limit, we get $M_\mathrm{dyn}\lesssim2.2\times10^{10}\,M_\odot$. However, if we assume \source\ to have similar size upper limits of $r_\mathrm{e}\approx200$\,pc as its high-redshift counterparts \citep{Akins2024, Kokorev2024LRD}, the estimated dynamical mass decreases to $M_\mathrm{dyn}\lesssim8.7\times10^8\,M_\odot$ --- similar physical size have also been estimated for a $z\sim2$ LRD with \jwst/NIRCam imaging by \citealt{Juodzbalis2024rossetta}). If the latter is true, then \source\ would also be the host of an overmassive black hole with $M_\mathrm{BH}/M_\star \approx 7\%$. One should note that although the narrow line width could trace the host's dynamical mass, the potentially high gas density could lead the \oiii\ emissions to be located farther out from the NLR \citep{Baggen2024}. Whether the narrow line width and the physical size traces the same spatial scale, and thus the dynamical mass, is also uncertain. 

As discussed above, \source\ shares many similarities with the high-redshift LRDs discovered by \jwst, including the presence of narrow \ha\ absorption, high Eddington ratio, and small dynamical masses. Along with the V-shaped photometric SED with a NIR flattening, we conclude that \source\ is indeed a LRD at cosmic noon and that finding them is feasible from the ground.


 

\section{Discussion and Summary}\label{sec:summary}
In summary, by surveying an area nearly 17 times larger than that of the previous LRD search within \jwst\ deep fields, we photometrically select LRD candidates at $1.7<z<3.7$ in HSC--SSP UD fields. We also spectroscopically confirm one of the LRD candidates at $z\approx2.2$ as our first target of the follow-up campaign. More importantly, we find that the number density of LRDs at cosmic noon indeed drops dramatically from $\sim10^{-4}\,\mathrm{cMpc^{-3}}$ at $z>4$ to $\lesssim3\times10^{-5}\,\mathrm{cMpc^{-3}}$ at $z<4$, a result that is qualitatively consistent with that of \cite{Kocevski2024}. Such a steep drop in LRD number density at $z=4$ could arise from changes related to the accretion of the central black hole.

Given the opposite number density evolution between the LRDs and UV-selected quasars shown in Figure~\ref{fig:zevolution}, we infer that LRDs represent a form of black hole growth only possible in the early universe due to super-Eddington accretion, high gas density, and/or low metallicity \citep{Inayoshi2025lognormal}. Additionally, the redshift evolution of massive galaxies also goes in the opposite direction to that of LRDs. This could serve as an indirect hint that LRDs are likely not galaxy-dominated systems, along with the challenges in SED fitting \citep[e.g.,][]{deGraaff2025bhstar, Ma2025Error404}. Qualitatively similar to LRDs, although possibly coincidental and at lower redshifts than LRDs are, the number density of compact galaxies also increases to $z\sim2$ from high redshifts before it drops by $\sim1\,\mathrm{dex}$ down to $z\sim0.5$ due to size changes driven by minor mergers \citep[e.g.,][]{Cassata2013masssize, vanderWel2014masssize}. 

Our number density estimation in Figure~\ref{fig:zevolution} agrees with the measurements by \cite{Inayoshi2025lognormal} from \jwst\ surveys at $z\sim3$. However, in the $z\sim2$ bin, our estimation is $\sim1\,\mathrm{dex}$ higher than the \jwst\ studies. It is challenging to obtain a larger sample and thus robust statistics for these rare objects at cosmic noon with the small survey volume covered by \jwst, which could account for this difference. 
Alternatively, if the contamination rate is as high as $\sim90\%$ in our sample, then the discrepancy may be reconciled. However, \source, being the first and successful follow-up we obtain, could be used to do a crude order-of-magnitude estimation for the LRD fraction $f_\mathrm{LRD}$. From Bayes theorem, we write $P(f_\text{LRD}|\text{1st one is LRD})\propto P(\text{1st one is LRD}|f_\text{LRD})P(f_\mathrm{LRD}) \propto f_\mathrm{LRD}$, where we assume a flat prior. This posterior has a fairly large $1\sigma$ confidence interval of 30\%--90\%. If this reflects the true contamination rate, it is possible that a different parameterization from a log-normal distribution \citep{Inayoshi2025break} may be required to describe the redshift evolution of LRDs, in addition to the the systematics described before. 

Indeed, our ground-based search suffers more from contaminating sources than the \jwst\ studies do. The limited angular resolution from the ground could also select ULIRGs with sufficiently small sizes. Our ongoing spectroscopic follow-up campaign will characterize the contamination rate. Deep space-based imaging surveys done by facilities with high NIR angular resolution such as \textit{Euclid} \citep{Euclid2024cosmicdawn} could also efficiently resolve this issue. Indeed, \cite{Bisigello2025euclidLRD} carry out a search of LRDs at $z\sim2$--3 with the \textit{Euclid} Quick Data Release 1. In contrast to our work, the authors find a number density that is roughly 10 times higher than our results. As the authors note, the Euclid sample likely suffers a higher contamination rate because of less robust SED shape constraints from measuring slopes on shallow optical photometry. 

Lastly, while our study has demonstrated the power of deep, large-area ground-based surveys to identify these rare objects at low-redshift and estimate their number densities, this sample, once spectroscopically followed-up, also offers opportunities to build multi-wavelengths SEDs from X-rays to radio like \cite{Akins2024} do for the high-$z$ sources. Leveraging the lower redshift, \jwst/MIRI observations could also be involved to constrain the temperature of warm dust, if any, in these systems, which is not feasible for the high-redshift counterparts \citep{Wang2024BRD, Setton2025ALMA}.

\section*{Acknowledgments}

Y. M. thanks Zhengrong Li, Xiaowei Ou, Sihao Cheng, and Rodrigo C\'ordova Rosado for useful discussion. Y. M. also thanks for the suggestions from Kohei Inayoshi. M. A. acknowledges support by the National Aeronautics and Space Administration (NASA) through an award (RSA 1628138) issued by JPL/Caltech. D. M., L. R., and A. S. acknowledge support by NASA under award number 80NSSC21K0630, issued through the Astrophysics Data Analysis Program (ADAP).

The Hyper Suprime-Cam (HSC) collaboration includes the astronomical communities of Japan and Taiwan, and Princeton University.  The HSC instrumentation and software were developed by the National Astronomical Observatory of Japan (NAOJ), the Kavli Institute for the Physics and Mathematics of the Universe (Kavli IPMU), the University of Tokyo, the High Energy Accelerator Research Organization (KEK), the Academia Sinica Institute for Astronomy and Astrophysics in Taiwan (ASIAA), and Princeton University.  Funding was contributed by the FIRST program from the Japanese Cabinet Office, the Ministry of Education, Culture, Sports, Science and Technology (MEXT), the Japan Society for the Promotion of Science (JSPS), Japan Science and Technology Agency (JST), the Toray Science Foundation, NAOJ, Kavli IPMU, KEK, ASIAA, and Princeton University.

This paper is based in part on data collected at the Subaru Telescope and retrieved from the HSC data archive system, which is operated by Subaru Telescope and Astronomy Data Center (ADC) at NAOJ. Data analysis was in part carried out with the cooperation of Center for Computational Astrophysics (CfCA) at NAOJ.  We are honored and grateful for the opportunity of observing the Universe from Mauna Kea, which has the cultural, historical and natural significance in Hawai`i.

This paper makes use of software developed for Vera C. Rubin Observatory \citep{Juric2017LSSTdatamanagement, Bosch2019lsstpipe, Ivezic2019lsst}. We thank the Rubin Observatory for making their code available as free software at \href{http://pipelines.lsst.io/}{http://pipelines.lsst.io/}. 

This paper also partially makes use of the CLAUDS data \citep{Sawicki2019CLAUDS}, whose data products can be accessed from \href{https://www.clauds.net}{CLAUDS data products can be accessed from https://www.clauds.net}.

\restartappendixnumbering
\appendix\label{sec:Appendix}
\section{Full Photometric Sample}\label{app:sample}
\begin{figure}[!ht]
    \centering
    \includegraphics[width=\textwidth]{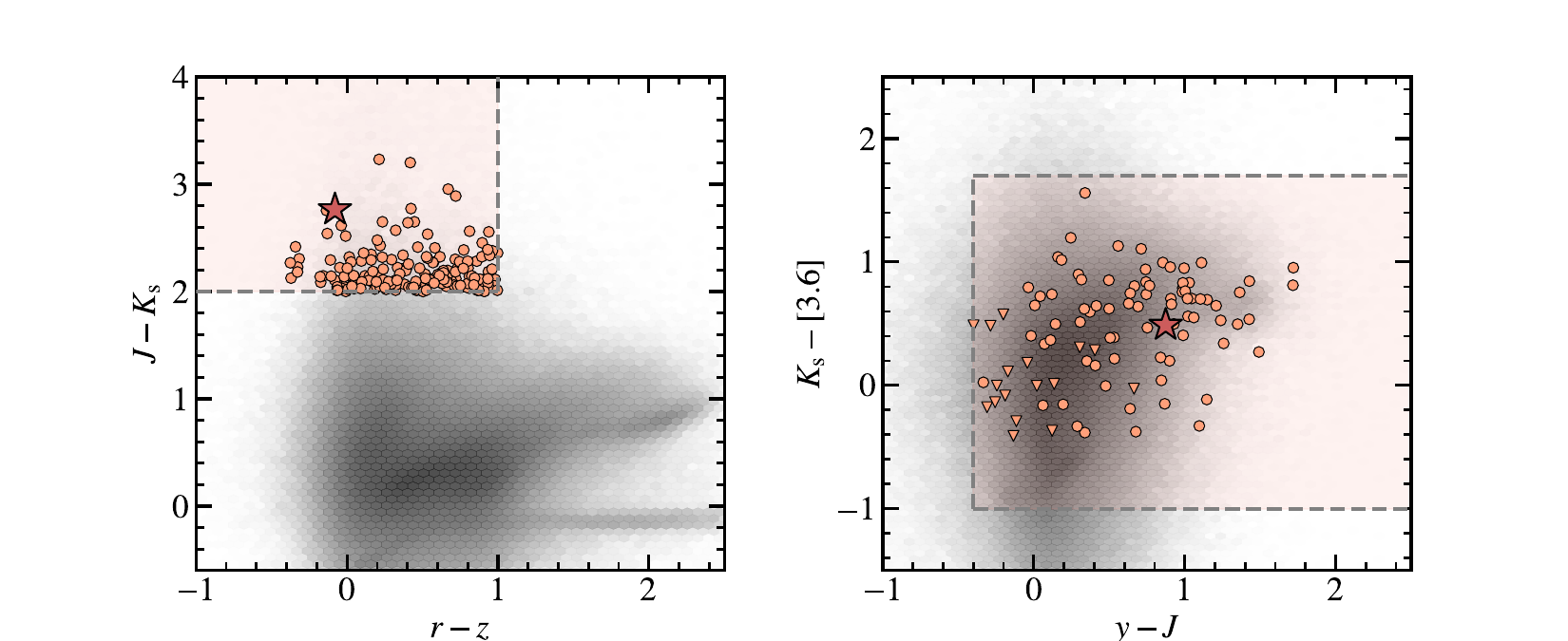}
    \caption{The sample of photometrically selected LRD candidates are shown in the color-color space as orange circles. Triangles represent upper limits. Shaded regions are the proposed LRD selection boxes. The hexagons in the background are all objects in the $U$--to--IRAC catalogs of the HSC-UltraDeep fields. The star represent our first spectroscopically followed-up candidate, \source. }
    \label{fig:sample}
\end{figure}

\begin{figure}
    \centering
    \includegraphics[width=0.6\textwidth]{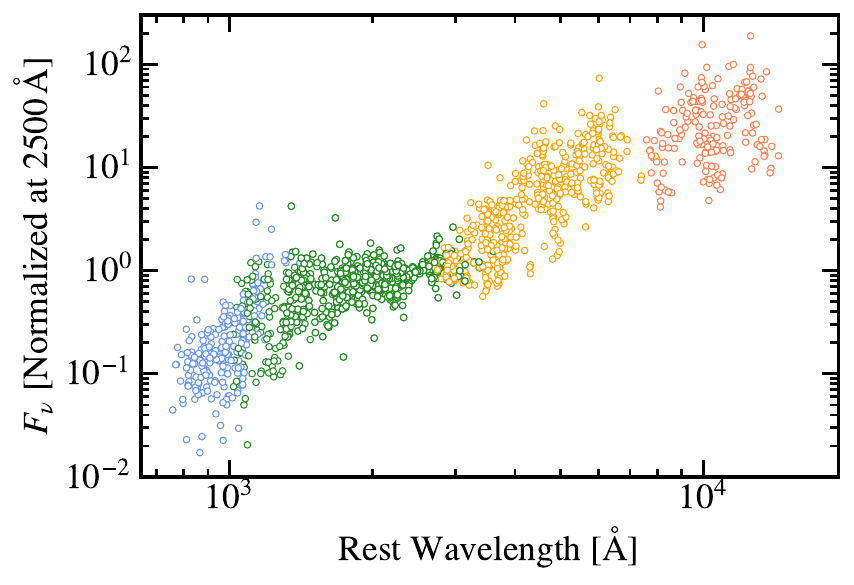}
    \caption{The photometric SEDs of the identified LRD candidates (normalized to 2500\,\AA) shown in the rest frame according to their photometric redshift estimations. Blue, gree, orange, and red colors represent photometry from CFHT/MegaCam, Subaru/HSC, VISTA/VIRCam, and \textit{Spitzer}/IRAC, respectively. For bands with $\mathrm{SNR}<3$, $3\sigma$-upper limits are shown. }
    \label{fig:sample_sed}
\end{figure}

The sample of LRD candidates we identified in the HSC-UD fields are shown in Table~\ref{tab:sample}. These candidates are used to estimate the number densities at $2\lesssim z\lesssim 4$ in Section~\ref{sec:number_density}. We also show these LRD candidates in the proposed color-color selection space in Figure~\ref{fig:sample} as well as their SEDs in Figure~\ref{fig:sample_sed}.

\begin{deluxetable}{cccccccc}[!ht]
\tablecaption{The sample of LRD candidates in the HSC-UltraDeep fields}
\tablewidth{0pt}
\tablehead{
\colhead{ID} & \colhead{RA} & \colhead{Dec} & \colhead{$z_\mathrm{phot}$} & \colhead{$m_J$} & \colhead{$m_{K_\mathrm{s}}$} & \colhead{$m_r$} & \colhead{$m_z$}\\ 
\colhead{} & \colhead{[deg]} & \colhead{[deg]} & \colhead{} & \colhead{[mag]} & \colhead{[mag]} & \colhead{[mag]} & \colhead{[mag]}
}
\startdata
37484705032989900 & 34.7984 & $-5.2526$ & 4.13 & 24.12 & 21.17 & 26.98 & 26.31 \\
43158876522044216 & 149.8855 & $+2.3728$ & 3.41 & 23.43 & 21.39 & 25.35 & 24.55 \\
37485130234751578 & 34.3009 & $-4.6948$ & 2.34 & 24.75 & 21.98 & 25.52 & 25.10 \\
37484700738020091 & 34.7997 & $-5.4402$ & 2.92 & 24.24 & 22.17 & 25.48 & 25.21 \\
43153799870700468 & 149.4046 & $+2.7348$ & 2.32 & 24.67 & 22.30 & 26.72 & 25.75 \\
43158859342178868 & 149.8604 & $+1.6461$ & 2.98 & 25.67 & 22.47 & 26.36 & 25.94 \\
37484979910915093 & 34.4148 & $-5.1374$ & 2.39 & 25.36 & 22.47 & 26.28 & 25.56 \\
43158468500147285 & 150.4671 & $+2.5320$ & 2.66 & 24.53 & 22.47 & 25.59 & 24.88 \\
43153645251890608 & 149.4681 & $+2.0947$ & 2.37 & 24.55 & 22.47 & 26.89 & 26.09 \\
43158455615244500 & 150.4510 & $+1.9663$ & 2.84 & 24.54 & 22.48 & 26.56 & 25.68 \\
\vdots & \vdots & \vdots & \vdots & \vdots & \vdots & \vdots & \vdots \\ 
\enddata
\tablecomments{We only show the first ten objects here. The full list will be available online. }
\label{tab:sample}
\end{deluxetable}

\section{Sample Contamination}\label{app:contamination}
\begin{figure}[!ht]
    \centering
    \includegraphics[width=0.9\textwidth]{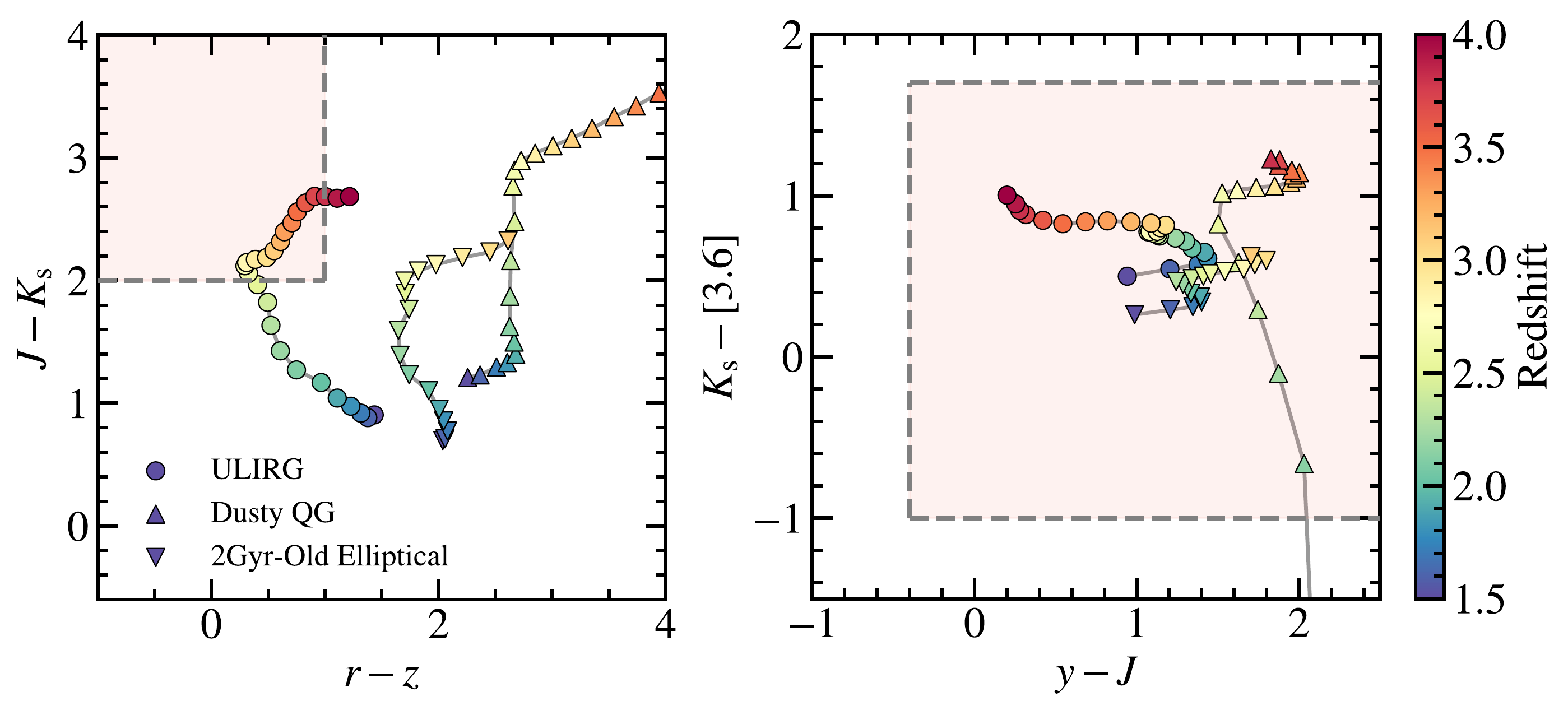}
    \caption{Redshift tracks of a ULIRG (Arp\,220), a dusty quiescent galaxy (dusty QG), and a 2\,Gyr-old elliptical galaxy on the color-color space. The dusty QG colors are computed based on the object in \cite{Setton2024dustyqg}. The other two templates are taken from the SWIRE template library \citep{Polletta2007swirelibrary}. The shaded red regions are the proposed color selection boxes for our LRD candidates.}
    \label{fig:contamination}
\end{figure}

We show in Figure~\ref{fig:contamination} the tracks of several other rest-frame optically red objects at cosmic noon in the proposed color space. ULIRGs are the major contaminants for our sample. 

\bibliography{ref}{}
\bibliographystyle{aasjournal}

\end{CJK*}
\end{document}